\documentclass[preprint,11pt]{elsarticle}
\biboptions{numbers,sort&compress}
\usepackage{url}
\usepackage{amsmath}
\usepackage{booktabs}

\begin{document}



\title{Bridging the Divide: Gender, Diversity, and Inclusion Gaps in Data Science and Artificial Intelligence Across Academia and Industry in the majority and minority worlds}
\author{Genoveva Vargas-Solar$^{1\dagger}$}
	\address[1]{CNRS, Univ Lyon,  INSA Lyon, UCBL, LIRIS, UMR5205, F-69221, France}
	%
	\address{genoveva.vargas-solar@cnrs.fr}



\begin{abstract} 
As Artificial Intelligence (AI) and Data Science (DS) become pervasive, addressing gender disparities and diversity gaps in their workforce is urgent. These rapidly evolving fields have been further impacted by the COVID-19 pandemic, which disproportionately affected women and minorities, exposing deep-seated inequalities. Both academia and industry shape these disciplines, making mapping disparities across sectors, occupations, and skill levels essential.
The dominance of men in AI and DS reinforces gender biases in machine learning systems, creating a feedback loop of inequality. This imbalance is a matter of social and economic justice and an ethical challenge, demanding value-driven diversity. Root causes include unequal access to education, disparities in academic programs, limited government investments, and underrepresented communities' perceptions of elite opportunities.
This chapter examines the participation of women and minorities in the fields of AI and DS, with a focus on their representation in both industry and academia. Analysing the existing dynamics seeks to uncover the collective and individual impacts on the lives of women and minority groups within these fields. Additionally, the chapter aims to propose actionable strategies to promote equity, diversity, and inclusion (DEI), fostering a more representative and supportive environment for all.

\end{abstract}


\maketitle



\section{Introduction} 
As promising domains in Information Technologies (IT), such as artificial intelligence (AI) and data science (DS), become increasingly ubiquitous, addressing gender disparities and diversity, equity and inclusion (DEI) gaps in the workforce of these disciplines is urgent and necessary. 
The rapid evolution of AI and DS, coupled with the disproportionate impact of the COVID-19 pandemic on women and minorities, has exacerbated existing inequalities, further exposing systemic challenges   \cite{berman2015,EC2019,EC2020a,EC2020b}. Industry and academia play an essential role in shaping these fields, making it critical to map how these disparities manifest across different sectors, occupations, and skill levels. This underscores the urgent need to critically examine these issues and implement strategies to mitigate disparities and foster a more inclusive workforce.

The persistent dominance of men in AI and DS, especially across various regions, contributes to a feedback loop that perpetuates gender biases within machine learning and AI systems. This unbalanced distribution of opportunities is fundamentally an issue of social and economic justice and an ethical challenge, highlighting the need for value-driven diversity. Understanding the root causes of this discriminatory landscape requires examining factors such as unequal access to education, disparities in the availability of undergraduate and graduate programs in these disciplines, variations in government investments in technology and education, and the perceptions underrepresented communities have about accessing elite opportunities.

 This chapter examines the participation of women and other minorities in IT like DS and AI within industry and academia across majority and minority contexts to explore their individual and collective impacts on women's lives.
Accordingly, the chapter is organised as follows. Section \ref{sec:women-in-it}  explores gender representation in AI and DS, comparing workforce participation across different countries, particularly within the European context, and analysing structural barriers such as mentorship access, funding, and cultural biases. It examines the historical biases embedded in AI and DS, the influence of societal perceptions on career choices, career progression obstacles, and the exacerbation of these challenges due to the COVID-19 pandemic. Section \ref{sec:curated-datasets} highlights the importance of gathering diverse workforce data and proposing various collection methods, such as surveys and industry reports, while addressing ethical considerations, including privacy and bias mitigation. Sections \ref{sec:analytics} and  \ref{sec:use-case} introduce and illustrate key workforce metrics such as turnover and attrition rates, comparing industry and academia and evaluating role-specific trends in managerial, technical, and research positions. It proposes solutions such as educational initiatives, workplace interventions like mentorship programs, inclusive hiring policies, and recommendations for industry-academia collaboration to foster greater diversity. Finally, section \ref{sec:conclusion} summarises the paper's key findings, suggests directions for future research, and calls for concerted efforts from stakeholders to reduce gender disparities in AI and DS, ultimately fostering a more inclusive workforce.

\section{Women in economically promising IT domains}
\label{sec:women-in-it}
Given the near-ubiquitous presence of AI and DS in modern society, addressing gender disparities and gaps in DEI within these fields has become imperative.  We focus on the French and European contexts to discuss the participation of women in economically promising IT domains.

France presents a unique case study as one of the leading European economies. The country has a rich technological ecosystem supported by government initiatives and an intense academic landscape. However, gender disparities persist, particularly in economically promising fields such as AI, DS, and other STEM-related industries. Despite governmental efforts to enhance gender diversity, women remain underrepresented in these fields, both in higher education and the workforce.
The fields are particularly fast-moving both in industry and academia, so it is essential to comprehensively map how these gaps   \cite{alfrey2017,DavenportPatil2012} are manifest across different sectors, occupations, and skills. While women represent nearly half of university students in France, their participation in AI and DS remains significantly low. Reports indicate that only around 15\%  of AI specialists in France are women, reflecting a stark gender imbalance. Structural barriers, including cultural biases, lack of mentorship, and limited access to funding and networking opportunities, contribute to this disparity.

Compared to other European countries, France falls behind leaders such as Sweden and Finland, where gender diversity initiatives in STEM fields have seen relative success \cite{EC2019}. These countries have implemented gender quotas, funding for female-led tech startups, and dedicated programs to encourage women's participation in IT-related disciplines. Conversely, countries like Germany and Italy also struggle with low female representation in IT, particularly in higher academia and research positions, mirroring the challenges observed in France \cite{EC2020b}. The representation of women in AI-related fields is particularly low in Southern and Eastern Europe, with countries such as Greece and Poland reporting some of the widest gender gaps in IT employment   \cite{ElementAI2019}.
%
However, women in France have been more likely to face job losses in tech-related fields due to the pre-existing gender gap in digital skills. The pandemic also intensified domestic and caregiving responsibilities, limiting women's ability to engage in reskilling programs or advance their careers in high-tech industries.

This ubiquitous male dominance   \cite{bolukbasi2016,cheryan2020,criadoperez2019} in countries in different regions results in a feedback loop shaping gender bias in AI and machine learning systems  \cite{Eubanks2018,Faulkner2009,YoungWajcmanSprejer2021}. AI algorithms are often trained on datasets that reflect historical gender biases, reinforcing and perpetuating systemic inequities. The underrepresentation of women in AI development means critical perspectives are missing to challenge and correct these biases. Without intentional intervention, the gender gap in AI and DS will continue to grow, ultimately affecting the fairness and Inclusivity of technological advancements.

We believe that the phenomenon should be studied by  country and then scaled to more global studies. In France, policies that promote gender-inclusive education, equitable hiring practices, and targeted mentorship programs are necessary to bridge the gap. Encouraging women's participation in AI and DS is an ethical and economic imperative—diverse teams drive innovation and create more robust AI models that serve a broader spectrum of society. Addressing these challenges will ensure France remains competitive in the global AI landscape while fostering an inclusive and fair technological future.

\subsection{AI and DS female workforce}

The persistent absence of women in AI and DS fields is troubling. According to the report of the World Economic Forum in 2018, over three-quarters of professionals in these fields globally are male (78\%); less than a quarter are women (22\%)   \cite{bobbittzeher2011}\footnote{World Economic Forum. Global Gender Gap Report 2018. Retrieved from: \url{https://www.weforum.org/reports/the-global-gender-gap-report-2018}.}. What about other underrepresented communities  \cite{atomico2020,benjamin2019}? How are they represented in the DS and AI workforces, and what are some of the opportunities offered by these promising areas taken by these communities? Of course, to acquire a full understanding of this phenomenon, it is necessary to treat the female community  \cite{Duke2018,UNWomen_WomenCount,HillCorbettStRose2010}. Other underrepresented communities  \cite{benjamin2019,Herring2009} are multifaceted and heterogeneous groups with a plurality of experiences, where gender intersects with multiple aspects of difference and disadvantage  \cite{collins1998}. 

Different academic organisations like IEEE (Women in Engineering), ACM (women in ACM ) in the US, the European Institute of Gender Equality in Europe, the CNRS (Centre National de la Recherche Scientifique) through the "Mission pour la Place des Femmes", and the "Comité parité-égalité", the Institute of Gender in France, and major technology companies (e.g., Microsoft, Google\footnote{	Google (2020). Google Diversity Annual Report 2020. Google.}, Facebook) have recognised the importance of understanding and organising actions that can promote a more gender-
balanced, diverse and inclusive STEM (Science, Technology, Engineering and Mathematics) community. The opportunities, economic share, and return on investment in different disciplines in STEM are not homogeneous. In the last decade, the role of data and the scientific and technical skills used to exploit it have created promising careers and economic spaces. Artificial Intelligence (AI) and Data Science (DS) have emerged as promising areas for developing careers with critical economic benefit perspectives. Nevertheless, professional career perspectives in these disciplines are not equal depending on gender \footnote{Contributes to Research and Innovation. Policy Review. H2020 Expert Group will update and expand 'Gendered Innovations/ Innovation through Gender'. Luxembourg: Publications Office of the European Union.}  \cite{WajcmanYoungFitzMaurice2020,WestWhittakerCrawford2019} and other criteria  \cite{FreirePorcaroGomez2021}, including race/ethnicity, socioeconomic level, the institution's reputation where people were educated, etc. For example, the House of Lords Select Committee on Artificial Intelligence in the United Kingdom 2018 advocated increasing gender and ethnic diversity among AI developers. In France, several companies like Renault and Engie, through the "Laboratoire de l'Egalité", signed a call for widespread awareness of the discriminatory effects of AI and a commitment by its supporters to correct them. It is addressed to leaders in the public and private sectors, research and training organisations, companies that produce digital technology, companies that use digital technology and AI consultants. In 2020, the European Commission \cite{EC2019,EC2020a,EC2020b} noted that it is 'high time to reflect precisely on the interplay between AI and gender equality'  \cite{EC2019}. French, European \cite{atomico2020}, and international organisations and agencies   \cite{abbate2012,alegria2019,ashcraft2016} perform studies for observing workforce shares in industry and sometimes in academia in a global perspective, often observing STEM \cite{QuirosEtAl2018}. Few fine grain studies have studied underrepresented communities' workforce evolution and gaps from the gender perspective in promising disciplines like AI and DS  \cite{best2019,ManthaHudson2018,UNESCO2020,YoungWajcmanSprejer2021}. A thorough understanding of the way the workforce accesses the AI and DS opportunities in industry  \cite{burtch2018,cardador2018} and academia   \cite{Ensmenger2012,FouldsIslamKeyaPan2019} is essential for building fair and inclusive societies  \cite{Simonite2018,AlanTuringInstitute_nodate}, but also for making sure that countries obtain full benefits of the development of these areas to achieve better economic and social conditions and leading positions in the international arena through technology self-sufficiency.

\subsection{Challenges} 
Despite the economic and symbolic capital \footnote{Symbolic capital, according to Pierre Bourdieu, is the prestige, recognition, and legitimacy an individual or group holds within a social field. It is a misrecognized form of other types of capital (economic, cultural, or social), gaining value through collective acknowledgment and legitimization. Symbolic capital grants authority and influence without direct coercion, shaping power dynamics in areas like academia, politics, and the arts.} investment seeking a fair distribution of AI and DS opportunities for women and underrepresented communities, the crystal ceiling must still be reached and broken  \cite{DobbinKalev2016}. Part of the explanation resides in data (!) \cite{buolamwini2018,data2x}. Indeed, many actors and studies agree that quality, disaggregated, intersectional data are still missing, and these data are essential to interrogate and tackle inequities in the AI and data science labour force [62]. As stated in the Alan Turing Institute study, "Where are women?"  \cite{YoungWajcmanSprejer2021}, the Royal Society has noted that a significant barrier to diversity is the lack of access to data on diversity statistics. The AI Roadmap recognises diversity and inclusion as priorities in making data-centred decisions to determine where to invest and ensure that underrepresented groups are given equal opportunity. 

We believe it is necessary to promote actions by data experts applying mathematical and machine learning techniques for building complete, high-quality, privacy-preserving data collections that can be used for performing studies that can drive conclusions about gender and DEI gaps in these disciplines.

To achieve a comprehensive and representative understanding of the dynamics shaping the status and perspectives of women in economically promising IT domains, it is essential to integrate both social and quantitative dimensions. This includes examining societal factors, such as patriarchal perceptions of gender roles, alongside measurable data. Collecting relevant data and establishing robust metrics to assess gender and qualification gaps within IT fields is crucial. These metrics should be systematically incorporated into curation and analytics pipelines, enabling ongoing monitoring and informed decision-making to address disparities and promote Inclusivity.

\section{Towards a curated and intersectional data collection } \label{sec:curated-datasets}
Understanding the professional landscape of women in AI, DS, Blockchain, and other economically promising IT expertise areas requires a robust data curation approach. The process involves systematically harvesting, integrating, archiving, and curating diverse digital and non-digital data sources that may not always be directly available. The primary goal is to construct comprehensive datasets that accurately represent the composition of the workforce, career progression, and opportunities while considering variables such as gender, ethnicity, socioeconomic status, and sexual orientation  \cite{smith2023datacuration,jones2022genderbias}.
We consider that there are two main research questions introduced by the challenge of curating datasets considering intersectional \footnote{Intersectionality is a sociological framework that provides a nuanced lens for analysing how overlapping social and political identities—such as gender, race, ethnicity, class, sexuality, religion, disability, age, caste, physical appearance, height, and weight—intersect to create distinct experiences of discrimination and privilege for individuals and groups. This approach highlights the complexity of systemic inequalities by emphasising how multiple dimensions of identity interact to shape unique lived realities \url{https://en.wikipedia.org/wiki/Intersectionality}. } aspects:
\begin{itemize}
    \item RQ1: How to build, curate and maintain an intersectional data collection integrating data from available sources about workforce in AI and DS respecting privacy and General Data Protection Regulation (GDPR)?
    \item RQ2: How to compute a quantitative profile of the gender and diversity and inclusion gap in AI and DS professional careers according to a methodology that can lead to the enlightening interpretation of results?
\end{itemize}

\subsection{Existing Data}
 The evidence base about gender diversity in the AI and DS workforce is minimal  \cite{barsan2020}. The available data is fragmented, incomplete and inadequate for investigating the career trajectories of women and men in the fields  \cite{data2x}. Available datasets often rely upon data produced through proprietary analyses and methodologies. Governmental statistics lack detailed information about job titles and pay levels within ICT, computing, and technology. This partial vision of the workforce status is a significant barrier to examining the emerging hierarchy between data science, AI, and other subdomains. Besides, available data about the global AI and DS workforce is often aggregated and rarely broken down by age, race, geography, (dis)ability, sexual orientation, socioeconomic status, and gender. As stated by  \cite{Wajcman1991,Wajcman2010,YoungWajcmanSprejer2021}:
 \begin{quote}
      "This is particularly concerning since it is those at the intersections of multiple marginalised groups who are at the greatest risk of being discriminated against at work and by resulting AI bias". 
 \end{quote}

Estimating the Number of women working in AI and DS across various platforms and regions is challenging due to limited publicly available data. However, existing studies and reports provide some insights:
\begin{itemize}
    \item Women constitute approximately 22\% of AI professionals worldwide. This underrepresentation is evident across various regions and platforms \cite{WELT2020}.
    \item In Germany, women make up about 20,3\% of AI professionals, which is below the international average. This places Germany among the countries with lower female representation in AI within Europe  \cite{WELT2020}.
    \item The United States leads with approximately 442,000 AI professionals, followed by India with 213,000, the United Kingdom with 69,000, and Germany with 58,000. While specific gender breakdowns for these countries are not provided, the global trend of underrepresentation suggests similar patterns  \cite{WELT2020}.
    \item Platform-Specific Insights:
    \begin{itemize}
        \item  Research indicates gender biases in participation and knowledge sharing on Stack Overflow, a major platform for programmers. Women are underrepresented, which may affect their visibility and engagement in the community  \cite{TANDFONLINE2021}.
        \item While specific numbers are scarce, studies suggest that women are underrepresented on GitHub, a leading platform for code sharing and collaboration. This underrepresentation can impact collaborative opportunities and project contributions \cite{Springer2024}.
    \end{itemize}
\end{itemize}

Accurately quantifying the Number of women in AI and DS roles across platforms like LinkedIn, ORCID, Google Scholar, Kaggle, Stack Overflow, and GitHub is difficult due to several factors:
\begin{itemize}
    \item Self-Reported Data: Many platforms rely on users to self-report gender, leading to incomplete datasets.
    \item Privacy Concerns: Users may choose not to disclose personal information, making it hard to gather comprehensive statistics.
    \item Dynamic Nature: The rapidly evolving tech landscape means data can quickly become outdated.
\end{itemize}

Although precise statistics remain challenging, existing research consistently highlights a pronounced underrepresentation of women in AI and DS across diverse regions and platforms. Addressing this disparity is critical to advancing DEI within the tech industry  \cite{DavenportPatil2012,Simonite2018,AlanTuringInstitute_nodate}. Understanding the extent of women's participation in these fields, their economic contributions, and their share of the financial benefits is essential for creating a more equitable and inclusive workforce. This analysis sheds light on the current gaps and underscores the broader economic and social advantages of fostering gender parity in AI and DS.
\subsection{Intersectionality and Inclusion in Workforce Data Collection}

From a Social and Human Sciences (SHS) perspective, it is essential to define Intersectionality in the AI and DS workforce by incorporating DEI aspects. A feminist, inclusive perception of data must guide the selection and analysis of workforce trends \cite{lopez2021intersectionality,jones2022genderbias}. A crucial element of this approach involves defining inclusion/exclusion criteria for data sources to ensure an intersectional data collection that integrates various marginalised and underrepresented groups.

Harvesting such data presents additional challenges since official documents and open data from companies, governmental agencies, and academia may lack direct information on these variables. Thus, developing methodologies to extract implicit information and infer missing demographic insights from structured and unstructured data becomes critical \cite{smith2023datacuration}.

From an IT perspective, data integration must balance access to private and sensitive information (e.g., gender, race, salary, sexual orientation, and socioeconomic status) and individual privacy protection. Strategies for ethically managing sensitive data must adhere to regulations such as the GDPR \cite{brown2020gdpr}, ensuring that the collected datasets remain anonymised and comply with legal and ethical standards.

Furthermore, ensuring data reinforcement strategies—such as bias detection, missing data imputation, and real-time updates—requires semi-automated solutions that integrate governance policies with algorithmic Transparency and explainability. AI-driven data curation methods must also ensure traceability and accountability for the sources included in the dataset.

\subsubsection{Principles of Intersectional and Inclusive Data Collection}

Our  data collection approach is guided by the following:
\begin{itemize}
    \item \textbf{Intersectionality}: Recognising the overlapping influences of gender, race, ethnicity, disability, and socioeconomic background on career opportunities.
    \item \textbf{Inclusivity}: Ensuring all workforce demographics, including underrepresented groups, are represented.
    \item \textbf{Transparency}: Providing transparent methodologies and ethical guidelines for data collection.
    \item \textbf{Privacy and Consent}: Ensuring compliance with data protection laws (e.g., GDPR) and obtaining informed consent from participants.
\end{itemize}

We propose a mumultifacetedpproach to gather comprehensive workforce data:\\
\noindent{\em 1. Surveys and Self-Reporting}
\begin{itemize}
    \item Conduct voluntary, anonymous surveys covering demographic, educational, and career-related aspects.
    \item Use standardised questions across platforms to ensure consistency.
    \item Offer multilingual survey options to increase participation across diverse populations.
\end{itemize}

\noindent{\em 2. Integration with Professional Platforms}
\begin{itemize}
    \item Leverage data from LinkedIn, ORCID, Google Scholar, and industry recruitment platforms.
    \item Employ AI-driven parsing tools to analyse skill sets, job mobility, and pay disparities.
\end{itemize}

\noindent{\em 3. Employer and Institutional Reporting}
\begin{itemize}
    \item Collaborate with IT companies and academic institutions to obtain workforce diversity reports.
    \item Encourage organisations to report gender pay gap analyses, recruitment trends, and retention statistics.
\end{itemize}

\noindent{\em 4. Data from Government and Public Records}
\begin{itemize}
    \item Utilise labour market data from national statistics agencies.
    \item Compare workforce participation rates across different socioeconomic backgrounds.
\end{itemize}

\noindent{\em Key Metrics for Workforce Analysis}
To identify disparities and trends, data collection should focus on:
\begin{itemize}
    \item \textbf{Demographics}: Gender, ethnicity, disability status, and socioeconomic background.
    \item \textbf{Educational Background}: Degree levels, institutions attended, access to STEM programs.
    \item \textbf{Employment Trends}: Role types, promotion rates, wage gaps, attrition rates.
    \item \textbf{Skill Development}: Training participation, certification completion, mentorship access.
    \item \textbf{Workplace Environment}: Employee satisfaction, harassment reports, inclusion initiatives.
\end{itemize}

\noindent{\em Ethical Considerations}
Ensuring ethical workforce data collection requires:
\begin{itemize}
    \item \textbf{Informed Consent}: Participants must understand data use and opt-in voluntarily.
    \item \textbf{Anonymisation}: Personal identifiers should be removed to protect privacy.
    \item \textbf{Bias Mitigation}: Implement measures to ensure diverse representation in datasets.
    \item \textbf{Regulatory Compliance}: Align with legal frameworks such as GDPR, EEOC, and labour policies.
\end{itemize}

\subsection{Curation for designing intersectional women careers on IT dataset}
Our methodology involves collecting and analysing professional profiles and career trajectories from prominent AI and DS platforms, including LinkedIn, ORCID, Google Scholar, Kaggle, StackOverflow, and GitHub  \cite{VasilescuEtAl2015}. The primary goal is to establish an intersectional data framework that captures detailed insights into professional careers, such as positions, profiles, and responsibilities while tracking French professionals' evolution and representation across these platforms. By integrating data from professional social networks (LinkedIn, ORCID, Google Scholar) and technical platforms (Kaggle, StackOverflow, GitHub, and MS Scientific Graph), we aim to provide a comprehensive understanding of workforce dynamics and identify patterns of participation and advancement in AI and DS fields. This approach will enable a deeper exploration of DEI within these domains.

Regarding intersectional variables, according to the data provided by the profiles of these platforms, it will be possible to consider the geographic location of professional activities, companies, studies, topics, gender, AI and DS topics/projects, and active career duration. In some cases, it will be possible to oppose the country(ies) where a person studied, worked and where they originated. This information can give a hint on race/ethnicity. Regarding studies, the person studying in private/public institutions with a specific reputation (according to, for instance, the Shanghai ranking) can give hints on possible sociocultural levels or membership to elite intellectual communities.

The curation of an intersectional dataset on women's careers in IT is a critical step toward fostering DEI in AI and DS. By integrating ethical data collection practices and leveraging analytical tools, this initiative can provide valuable insights to contribute to a more inclusive technology landscape. Moving forward, continuous updates and interdisciplinary collaborations will be essential to ensure the dataset remains relevant and impactful.

\subsection{Data curation pipeline}
\begin{figure}[!b]\centering
	\includegraphics[width=0.95\textwidth
    ]{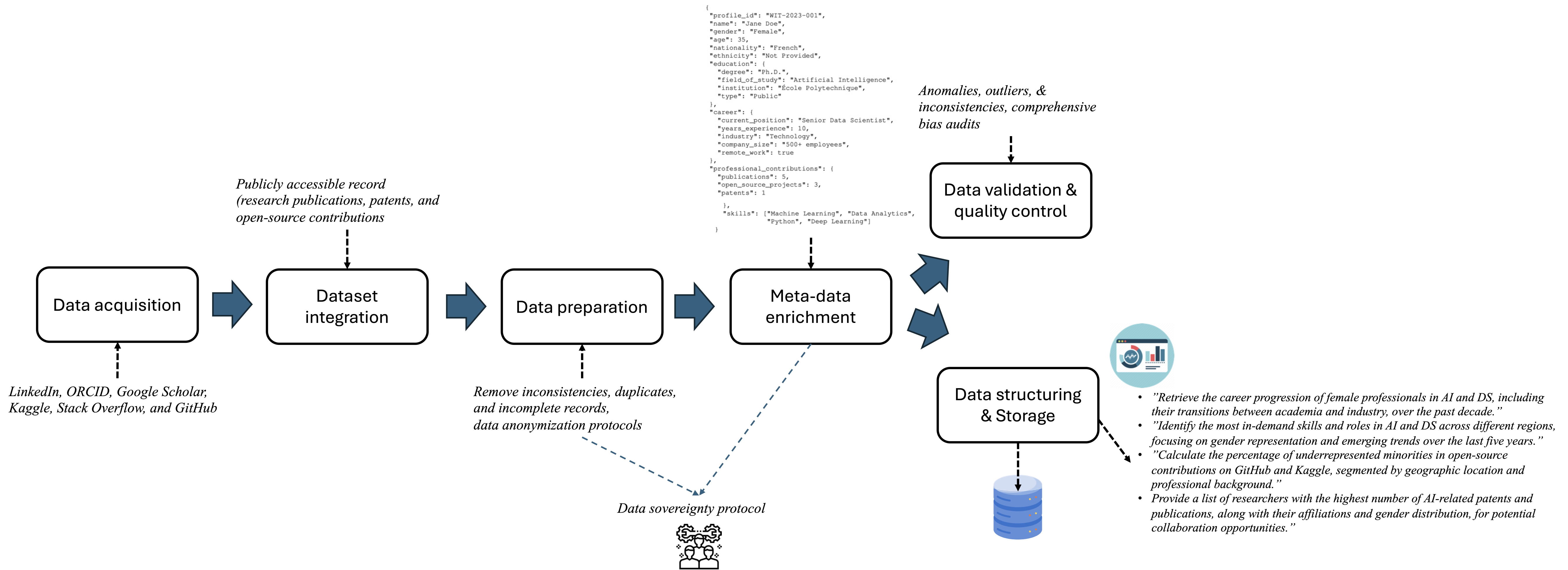}
	{\caption{Data curation pipeline\label{ch:fig01}}}
\end{figure}

The curation pipeline and metadata schema provide a comprehensive structure to understand the representation of women in IT. The data curation process follows a structured pipeline comprising six key stages (see Figure \ref{ch:fig01}):
\begin{enumerate}
    \item Data acquisition: To construct a comprehensive dataset, raw career data is gathered from various platforms, such as LinkedIn, ORCID, Google Scholar, Kaggle, Stack Overflow, and GitHub, to ensure a broad representation of professional and academic activities. Additionally, publicly accessible records—including research publications, patents, and open-source contributions—are systematically collected and integrated into the dataset. This multi-source approach ensures a robust and holistic foundation for analysis, enabling a thorough examination of career trajectories, contributions, and representation across the fields of AI and DS.

    \item Data Preprocessing: The collected data is standardised into a unified format to ensure consistency and compatibility across all platforms. This process involves cleaning and normalising entries to remove inconsistencies, duplicates, and incomplete records, thereby improving the overall quality and reliability of the dataset. Additionally, rigorous data anonymisation protocols are applied to protect sensitive information, ensuring full compliance with privacy regulations and ethical standards. This step is essential for creating a clean, secure, usable dataset ready for in-depth analysis.

    \item Metadata Enrichment: Enhance raw data by adding structured metadata fields, including:
\begin{itemize}
    \item Personal Attributes: Gender, age, nationality, ethnicity (where available and ethically collected).
    \item Education and Training: Degree levels, institutions attended, field of study, type of institution (public/private).
    \item Career Progression: Job titles, employment transitions, industry versus academia affiliations.
    \item Geographic Mobility: Countries of study, employment, and professional affiliations.
    \item Professional Contributions: Research publications, patents, repositories, collaborations, leadership roles.
    \item Skills and Expertise: Areas of specialisation such as AI, data science, cybersecurity, and software engineering.

Each entry in the dataset follows a structured format with metadata fields:
\begin{verbatim}
 {
  "profile_id": "WIT-2023-001",
  "name": "Jane Doe",
  "gender": "Female",
  "age": 35,
  "nationality": "French",
  "ethnicity": "Not Provided",
  "Education": {
    "degree": "Ph.D.",
    "field_of_study": "Artificial Intelligence",
    "institution": "École Polytechnique",
    "type": "Public"
  },
  "career": {
    "current_position": "Senior Data Scientist",
    "years_experience": 10,
    "industry": "Technology",
    "company_size": "500+ employees",
    "remote_work": true
  },
  "professional_contributions": {
    "publications": 5,
    "open_source_projects": 3,
    "patents": 1
  },
  "skills": ["Machine Learning", "Data Analytics", 
             "Python", "Deep Learning"]
}   
\end{verbatim}

\item Workplace Attributes: Company or organisation size, diversity policies, remote work options, work-life balance.
\end{itemize}
\item Data Validation and Quality Control: To ensure the accuracy and reliability of the dataset, cross-referencing is conducted with multiple independent sources to validate the integrity of the collected information. Advanced quality control algorithms detect anomalies, outliers, and inconsistencies, ensuring the dataset meets high-reliability standards. Furthermore, comprehensive bias audits are performed to identify and address potential underrepresentation or data skew, fostering fairness and representativeness in the analysis. These meticulous steps are critical to maintaining the dataset's credibility, robustness, and suitability for meaningful insights.

\item  Data Structuring and Storage: The curated data is systematically organised into a persistence system, enabling efficient storage, retrieval, and management. Metadata schemas are developed to enhance advanced querying capabilities, allowing users to explore the data by career trajectories, industry trends, and diversity metrics. API-based access is implemented to support collaboration and controlled data sharing, offering secure and scalable interfaces for authorised users to interact with the dataset. This structured approach ensures the data remains accessible, adaptable, and well-suited for various analytical and research applications. The following queries can be asked:
\begin{itemize}
    \item "Retrieve the career progression of female professionals in AI and DS, including their transitions between academia and industry, over the past decade."
    \item "Identify the most in-demand skills and roles in AI and DS across different regions, focusing on gender representation and emerging trends over the last five years."
    \item "Calculate the percentage of underrepresented minorities in open-source contributions on GitHub and Kaggle, segmented by geographic location and professional background."
    \item "Provide a list of researchers with the highest number of AI-related patents and publications, along with their affiliations and gender distribution, for potential collaboration opportunities."
\end{itemize}
\end{enumerate}

Curated datasets are developed and managed under strict ethical guidelines to prioritise privacy, fairness, and compliance with data protection laws. Informed consent is a cornerstone of data collection, ensuring all processes align with ethical research standards and respect participant autonomy. Rigorous data anonymisation techniques are applied to remove sensitive personal identifiers, safeguarding individual privacy. To promote fairness, bias mitigation strategies are implemented to prevent gender, racial, or other forms of discrimination in data representation. Additionally, the project adheres to regulatory frameworks such as GDPR and other data governance standards, ensuring legal and ethical integrity throughout the data lifecycle.

\section{Analytics on AI and DS gender and DEI gap} \label{sec:analytics}
Once the intersectional workforce dataset is curated, data analytics experiments can be designed to extract meaningful insights on the gender and DEI gaps in AI, DS, and IT-related fields. These experiments should focus on quantitative modelling and exploratory data analysis to answer key demographic and professional questions:
\begin{itemize}
    \item What is the distribution of women across AI and DS roles compared to men?
    \item How does gender representation vary across job positions, seniority levels, and domains (academia, industry, entrepreneurship)?
    \item What patterns emerge in salary gaps, career progression, and hiring trends across genders and intersectional variables?
    \item How do regional and economic factors influence the representation of women and minorities in IT expertise fields?
\end{itemize}

Implementing these experiments must leverage data mining, statistical analysis, and machine learning techniques to model inequalities, project future workforce trends, and propose policy interventions. This can also enable predictive modelling of gender representation in emerging IT fields and suggest interventions for increasing women's participation in high-impact technology areas such as blockchain development, cybersecurity, and quantum computing   \cite{jones2022genderbias,lopez2021intersectionality}.

\subsection{Quantitative methodology for measuring a DEI gap}
We propose a quantitative and analytical data science methodology to generate a first map of the gender, diversity, and inclusion gap in AI and DS professional careers.
Using the first intersectional data collection, we observe the following aspects that  lead to a gender and DEI map, applying a data science experiment using statistics and classification methods: 
\subsubsection{Diverging career trajectories} 
 of DS and AI professionals differentiated by gender  \cite{case2012,correll2001} (e.g., job segregation and skills specialisations). We also estimate the gender gap using proportions depending on the data we collect and the distribution by gender. We confront our findings with the Global Gender Gap report   \cite{schwab2017global}.
    Research indicates that women are more likely to occupy data preparation and exploration roles. At the same time, men are predominantly found in positions emphasising machine learning, big data, and computer science skills   \cite{case2012,correll2001}. This delineation suggests a structured gender inequality within these professions \cite{Wiley2021}.
Quantifying the gender gap in these fields reveals that women constitute approximately 22\% of AI professionals globally, leaving a significant disparity yet to be addressed \cite{WEF2018}. This underrepresentation is even more pronounced at senior levels, with women holding less than 14\% of executive roles in AI \cite{StiftungNV2021}.
These findings align with broader trends highlighted in the Global Gender Gap Report, which benchmarks gender parity across various dimensions, including economic participation and opportunity. The report underscores persistent gaps in labour force participation and leadership roles, reflecting systemic challenges that transcend individual industries \cite{WEF2024}.

\subsubsection{Self-reported skills} 
We identify skills reported in professional platforms/social networks and classify profiles concerning these skills   \cite{altenburger2017,roca2019identifying,tockey2019gender}, of course, classified by gender and academia vs industry. Analysing skills reported on professional platforms such as LinkedIn, ORCID, Google Scholar, and Microsoft Academic Graph reveals notable gender disparities across academia and industry.
\begin{itemize}
    \item LinkedIn: A study analysing global professional gender gaps using LinkedIn data found that women are significantly underrepresented relative to men on LinkedIn in countries in Africa, the Middle East, and South Asia, among older age groups, and in industries such as software and IT services, manufacturing, and finance \cite{Bursztyn2021}. Conversely, women are overrepresented in industries like healthcare and education. This suggests a gender-based distribution of skills, with men more prevalent in technical fields and women in caregiving and educational roles   \cite{Springer2021}.
    \item ORCID and Google Scholar: While specific data on gender differences in skills reported on ORCID and Google Scholar is limited, these platforms primarily showcase research outputs and affiliations. Studies have indicated that women in academia often have shorter publishing careers and higher dropout rates, leading to fewer publications and citations compared to their male counterparts \cite{Huang2020}. This disparity can influence the visibility of skills and expertise on these platforms.
    \item Microsoft Academic Graph: The Academia/Industry DynAmics (AIDA) Knowledge Graph, which integrates data from the Microsoft Academic Graph, provides insights into research dynamics across academia and industry. While it does not explicitly categorise skills by gender, it does highlight the distribution of research topics and industrial sectors   \cite{AIDA2021}. Integrating gender data could offer a more comprehensive understanding of industry skill disparities.
    \item Gender Disparities in Academia vs. Industry: Research indicates that women are underrepresented in technology entrepreneurship. An analysis of LinkedIn profiles revealed that females were only half as likely as males to find businesses in the technology industry \cite{Gompers2022}. This underrepresentation extends to leadership roles, with fewer women attaining executive positions in tech sectors \cite{INFORMS2022}.
\end{itemize}

Professional platforms reflect existing gender disparities in skill representation across academia and industry. Men are more prevalent in technical and leadership roles, while women are more represented in the educational and caregiving sectors. Addressing these disparities requires targeted interventions to promote equitable skill development and career advancement opportunities for all genders.

\subsubsection{Job turnover and attrition rates.} 
    We measure the turnover (i.e., changing job roles) and attrition rates (i.e., leaving the industry altogether) between women and men in the IT industry and then zoom in regarding DS and AI departments and R\&D groups   \cite{cech2011,dastin2018}. We also compare this situation in academia by observing research topics and the mobility of teaching majors.
To measure turnover rates (changing job roles) and attrition rates (leaving the industry altogether) between women and men in the IT industry, particularly in DS and AI departments and R\&D groups, we  define the following formulas:
\begin{itemize}
    \item {\em\bf Turnover Rate (TR)}
  Captures the proportion of employees who change job roles within a given period:
\begin{equation}
    TR = \frac{E_{\text{change}}}{E_{\text{total}}} \times 100
\end{equation}
where:
\begin{itemize}
    \item $E_{\text{change}}$ = Number of employees who switched roles within a given period.
    \item $E_{\text{total}}$ = Total number of employees at the beginning of the period.
\end{itemize}

A gender-specific turnover rate can be measured as follows:
\begin{equation}
    TR_W = \frac{E_{\text{change, women}}}{E_{\text{total, women}}} \times 100, \quad TR_M = \frac{E_{\text{change, men}}}{E_{\text{total, men}}} \times 100
\end{equation}

A \textbf{Turnover Gender Ratio (TGR)} is computed as:
\begin{equation}
    TGR = \frac{TR_W}{TR_M}
\end{equation}
where:
\begin{itemize}
    \item $TGR > 1$ suggests that women have higher turnover than men.
    \item $TGR < 1$ suggests men have higher turnover than women.
\end{itemize}

\item {\em Attrition Rate (AR)}
measures the proportion of employees who leave the industry altogether:
\begin{equation}
    AR = \frac{E_{\text{left}}}{E_{\text{total}}} \times 100
\end{equation}
where:
\begin{itemize}
    \item $E_{\text{left}}$ = Number of employees who left the industry.
    \item $E_{\text{total}}$ = Total number of employees at the beginning of the period.
\end{itemize}

Gender-specific attrition rates are defined as:
\begin{equation}
    AR_W = \frac{E_{\text{left, women}}}{E_{\text{total, women}}} \times 100, \quad AR_M = \frac{E_{\text{left, men}}}{E_{\text{total, men}}} \times 100
\end{equation}

A \textbf{Attrition Gender Ratio (AGR)} is computed as:
\begin{equation}
    AGR = \frac{AR_W}{AR_M}
\end{equation}
where:
\begin{itemize}
    \item $AGR > 1$ suggests that women have a higher attrition rate than men.
    \item $AGR < 1$ suggests men have a higher attrition rate than women.
\end{itemize}

\item {\em Zooming into DS \& AI Departments and R\&D Groups.}
We further analyse turnover and attrition by specific roles in the industry:
\begin{itemize}
    \item \textbf{Technical Roles}: AI engineers, ML scientists, data analysts.
    \item \textbf{Managerial Roles}: AI project leads, research heads.
    \item \textbf{Support Roles}: HR, administrative positions in IT companies.
\end{itemize}

We calculate the role-specific turnover and attrition rates:
\begin{equation}
    TR_{\text{R\&D}} \text{ vs. } TR_{\text{DS/AI}}
\end{equation}
\begin{equation}
    AR_{\text{R\&D}} \text{ vs. } AR_{\text{DS/AI}}
\end{equation}

\item {\em Comparison with Academia.}
We analyze \textbf{research topic mobility} and \textbf{teaching mobility} using:

\textbf{1. Changes in Research Focus (Topic Shift Rate - TSR)}:
\begin{equation}
    TSR = \frac{R_{\text{shift}}}{R_{\text{total}}} \times 100
\end{equation}
where:
\begin{itemize}
    \item $R_{\text{shift}}$ = Number of researchers shifting research areas.
    \item $R_{\text{total}}$ = Total number of researchers.
\end{itemize}

\textbf{2. Mobility of Teaching Majors (Teaching Mobility Rate - TMR)}:
\begin{equation}
    TMR = \frac{T_{\text{moved}}}{T_{\text{total}}} \times 100
\end{equation}
where:
\begin{itemize}
    \item $T_{\text{moved}}$ = Number of teachers who changed disciplines/universities.
    \item $T_{\text{total}}$ = Total number of teachers.
\end{itemize}
\end{itemize}
    
    \subsubsection{\bf Gender diversity in industry and academia and crystal ceiling phenomenon} 
We measure to which extent men and women are represented in AI and DS areas in industry and academia, observing the type of positions they achieve (i.e., the evolution of careers)   \cite{cech2011, WynnCorrell2017}. Are female professionals in AI and Data Science (DS) more likely to be concentrated in roles or duties that are traditionally feminised, such as student mentoring and tutoring, coordination of study programs, dissemination and communication tasks, or human resources? If so, how can this phenomenon inform discussions around legal rights for equal job access? Could this observation serve as a basis for highlighting and addressing grey areas in work-related legal rights, particularly in France and Europe? By examining these trends, we can better understand systemic biases and advocate for policies that ensure equitable opportunities and representation across all professional domains.

\subsubsection{\bf\em The qualification gap}
The qualification gap measures the achievement gap and defines clusters by types of skills associated with different AI and DS profiles (e.g., DS manager/engineer, AI practitioner or algorithms design). We use skills related to position profiles in the case of industry and academic positions in the case of academia, including coordination of education programs and leading positions.

To measure the \textbf{qualification gap} in AI and DS, we define methods to quantify the \textbf{achievement gap} and categorise professionals into clusters based on skills associated with different AI and DS profiles. These profiles vary by industry and academia, where other qualifications and skill sets are required.

\noindent{\em Defining the Qualification Gap.}
The \textbf{qualification gap (QG)} measures disparities in skills and competencies between AI and DS professionals. It can be defined as:
\begin{equation}
    QG = \frac{S_{\text{required}} - S_{\text{actual}}}{S_{\text{required}}} \times 100
\end{equation}
where:
\begin{itemize}
    \item $S_{\text{required}}$ = Skills required for a specific role (industry or academia).
    \item $S_{\text{actual}}$ = Skills possessed by individuals in that role.
\end{itemize}

A higher \textit{QG} indicates a more significant gap in necessary competencies.

\noindent{\bf Cluster Analysis Based on Skills.}
We classify professionals into clusters to analyse the qualification gap based on their job roles and associated skills.

In industry, AI and DS professionals are categorised as:
\begin{itemize}
    \item \textbf{Data Science Manager} (Project management, strategic AI decision-making, cross-functional leadership).
    \item \textbf{DS Engineer} (Data modelling, statistical analysis, big data management).
    \item \textbf{AI Practitioner} (Machine learning implementation, AI ethics, deep learning).
    \item \textbf{Algorithm Designer} (Algorithm development, computational efficiency, AI optimisation).
\end{itemize}

The skill set for each role is analysed by computing the following:
\begin{equation}
    S_{\text{Gap}} = \sum_{i=1}^{n} (S_{i, \text{required}} - S_{i, \text{actual}})
\end{equation}
where $n$ represents the Number of essential skills needed for the role.

Academic professionals in AI and DS are categorised as follows:
\begin{itemize}
    \item \textbf{Faculty Researcher} (Publications, research grants, innovation in AI methodologies).
    \item \textbf{Program Coordinator} (Academic administration, curriculum design, mentorship).
    \item \textbf{Department Head/Dean} (Leadership, institutional AI strategy, funding acquisition).
\end{itemize}

The qualification gap in academia is computed similarly but with additional emphasis on educational program management and research impact:
\begin{equation}
    QG_{\text{Academic}} = \frac{(P_{\text{required}} - P_{\text{actual}}) + (T_{\text{required}} - T_{\text{actual}})}{2} \times 100
\end{equation}
where:
\begin{itemize}
    \item $P$ = Publication output, citation impact, grant acquisition.
    \item $T$ = Teaching experience, mentorship, program coordination skills.
\end{itemize}

\noindent{\bf Comparative Reporting.}
To compare industry vs academia qualification gaps, we introduce a \textbf{standardised qualification gap measure}:
\begin{equation}
    SQG = \frac{QG_{\text{Industry}} + QG_{\text{Academic}}}{2}
\end{equation}
A higher \textit{SQG} value indicates a broader disparity in qualifications across AI and DS disciplines.

We can identify critical deficiencies in AI and DS expertise by quantifying qualification gaps and clustering professionals based on skills. This enables targeted policy and training interventions to bridge these gaps, ensuring alignment between education, industry demands, and research excellence.

\subsection{Understanding the dynamics of occupational segregation}
 The underrepresentation of women in AI and DS spans a wide range of roles, with minimal participation in technical and leadership positions. While specific data on the concentration of women in traditionally feminised roles—such as mentoring, program coordination, communication, or human resources—within AI and DS is scarce, broader labour market trends reveal that women are often overrepresented in these areas. Although these roles are critical to organisational success, they tend to be undervalued in compensation, recognition, and career advancement opportunities compared to technical or leadership roles. This disparity highlights the need to address systemic biases and re-evaluate how essential but non-technical roles are perceived and rewarded within the AI and DS ecosystem.

In France, the legal framework strongly supports gender equality in the workplace. The Labour Code mandates equal pay for men and women performing the same work or work of equal value   \cite{Leglobal2022}. Additionally, the Rixain Law introduces gender-based quotas and expanded reporting requirements to promote gender equality in the workplace \cite{WTWCO2022}.
However, the persistent clustering of women in traditionally feminised roles can obscure disparities in job access and advancement opportunities. This phenomenon highlights potential grey areas in the enforcement of equal employment rights. For instance, if women are systematically funnelled into specific roles, it may indicate indirect discrimination, even in the absence of overt bias. Addressing this issue requires a multi-faceted approach:

\begin{itemize}
    \item Comprehensive Data Collection: Gather detailed data on role distribution by gender within AI and DS to identify patterns of occupational segregation.

    \item Policy Enforcement: Ensure strict adherence to existing gender equality laws, focusing on equitable role distribution and advancement opportunities.

    \item Awareness and Training: Implement programs to raise awareness about unconscious biases that may influence role assignments and provide training to promote inclusive practices.

    \item Supportive Networks: Establish mentorship and sponsorship programs to encourage women to pursue and remain in technical and leadership roles within AI and DS.
\end{itemize}

By recognising and addressing the subtle dynamics contributing to occupational segregation, France and other European countries can strengthen the legal and practical frameworks that uphold equal job access, ensuring that gender equality extends beyond legislation into everyday workplace practices. 


\section{Scenario: Gender Disparities in AI and DS Career Mobility} \label{sec:use-case} 
This use case demonstrates the application of gender gap metrics in IT, specifically in DS and AI departments and R\&D groups, to measure turnover and attrition rates. These metrics are then compared with academia to identify gender disparities in career progression and mobility.

A multinational AI company conducts a study over one year to assess employee turnover and attrition rates in its Data Science, AI, and R\&D departments. The company had 500 employees at the start of the period, 200 women and 300 men.

\subsection{Turnover Rate (TR)}

The turnover rate measures the percentage of employees changing roles:
\begin{equation}
    TR = \frac{E_{\text{change}}}{E_{\text{total}}} \times 100
\end{equation}
where:
\begin{itemize}
    \item $E_{\text{change}} = 40$ employees changed roles (15 women, 25 men).
    \item $E_{\text{total}} = 500$ employees at the beginning of the period.
\end{itemize}

For gender-specific turnover rates:
\begin{equation}
    TR_W = \frac{15}{200} \times 100 = 7.5\%
\end{equation}
\begin{equation}
    TR_M = \frac{25}{300} \times 100 = 8.33\%
\end{equation}

The Turnover Gender Ratio (TGR) is calculated as:
\begin{equation}
    TGR = \frac{7.5}{8.33} = 0.9
\end{equation}
Since $TGR < 1$, men have a slightly higher turnover rate than women.

\subsection{Attrition Rate (AR)}

The attrition rate measures the percentage of employees leaving the industry:
\begin{equation}
    AR = \frac{E_{\text{left}}}{E_{\text{total}}} \times 100
\end{equation}
where:
\begin{itemize}
    \item $E_{\text{left}} = 30$ employees left the industry (18 women, 12 men).
    \item $E_{\text{total}} = 500$ employees at the beginning of the period.
\end{itemize}

For gender-specific attrition rates:
\begin{equation}
    AR_W = \frac{18}{200} \times 100 = 9\%
\end{equation}
\begin{equation}
    AR_M = \frac{12}{300} \times 100 = 4\%
\end{equation}

The Attrition Gender Ratio (AGR) is calculated as:
\begin{equation}
    AGR = \frac{9}{4} = 2.25
\end{equation}
Since $AGR > 1$, women have a significantly higher attrition rate than men.

\subsection{Comparison Across Roles}

To analyse DS, AI, and R\&D departments, turnover and attrition rates are computed for:
\begin{itemize}
    \item \textbf{Technical Roles}: AI engineers, ML scientists, data analysts.
    \item \textbf{Managerial Roles}: AI project leads, research heads.
    \item \textbf{Support Roles}: HR, administrative positions in IT companies.
\end{itemize}

\subsubsection{Comparison with Academia}
In academia, research and teaching mobility are measured using:
\subsection{Topic Shift Rate (TSR)}
\begin{equation}
    TSR = \frac{R_{\text{shift}}}{R_{\text{total}}} \times 100
\end{equation}
where:
\begin{itemize}
    \item $R_{\text{shift}} = 10$ researchers changed topics (7 women, three men).
    \item $R_{\text{total}} = 100$ researchers.
\end{itemize}

A Topic Shift Gender Ratio (TSGR) is calculated as:
\begin{equation}
    TSGR = \frac{TSR_W}{TSR_M} = \frac{\frac{7}{50} \times 100}{\frac{3}{50} \times 100} = \frac{14}{6} = 2.33
\end{equation}

\subsubsection{Teaching Mobility Rate (TMR)}

\begin{equation}
    TMR = \frac{T_{\text{moved}}}{T_{\text{total}}} \times 100
\end{equation}
where:
\begin{itemize}
    \item $T_{\text{moved}} = 8$ teachers moved institutions (5 women, 3 men).
    \item $T_{\text{total}} = 80$ teachers.
\end{itemize}

A Teaching Mobility Gender Ratio (TMGR) is calculated as:
\begin{equation}
    TMGR = \frac{TMR_W}{TMR_M} = \frac{\frac{5}{40} \times 100}{\frac{3}{40} \times 100} = \frac{12.5}{7.5} = 1.67
\end{equation}

\subsection{Discussion}
The  table \ref{tab:gender_gap_metrics} summarises the key gender gap metrics across industry and academia:

\begin{table}[h]
    \centering
    \begin{tabular}{lccc}
        \toprule
        \textbf{Metric} & \textbf{Women (\%) } & \textbf{Men (\%) } & \textbf{Gender Ratio} \\
        \midrule
        Turnover Rate (TR) & 7.5 & 8.33 & 0.90 \\
        Attrition Rate (AR) & 9.0 & 4.0 & 2.25 \\
        Topic Shift Rate (TSR) & 14.0 & 6.0 & 2.33 \\
        Teaching Mobility Rate (TMR) & 12.5 & 7.5 & 1.67 \\
        \bottomrule
    \end{tabular}
    \caption{Comparison of Gender-Based Turnover and Attrition Metrics in IT and Academia}
    \label{tab:gender_gap_metrics}
\end{table}

By applying gender-based turnover and attrition metrics, the company and academic institutions identified key trends:
\begin{itemize}
    \item Women exhibit lower turnover but higher attrition rates than men.
    \item Women in academia show more topic shifts and higher teaching mobility rates than men.
    \item Higher attrition among women suggests a need for targeted retention strategies in both industry and academia.
\end{itemize}
These insights enable organisations to develop more inclusive workforce policies and career support programs.

\section{Conclusions} \label{sec:conclusion}
This study lays the foundation for a data-driven approach to assessing gender and Intersectionality in artificial intelligence (AI) and data science (DS) by defining a structured data curation pipeline. This pipeline enables the systematic collection, standardisation, and analysis of professional and academic data from diverse sources, ensuring a comprehensive understanding of workforce dynamics. However, to build on this foundation and address existing gaps, future work should focus on scaling and enhancing these efforts through several key initiatives.

First, expanding data sources is critical to improving the representativeness and depth of the analysis. Additional datasets from industry reports, university records, and crowdsourced repositories will provide a more holistic view of workforce participation and career trajectories. This expansion will help capture underrepresented voices and ensure that the data reflects the full spectrum of professional experiences in AI and DS.

Second, enhancing AI-driven bias detection is essential to ensure the fairness and accuracy of demographic insights. By implementing fairness-aware algorithms, researchers can identify and mitigate biases within datasets, such as underrepresentation or skewed data distributions. This step is crucial for producing reliable and equitable analyses that inform meaningful interventions.

Third, developing interactive dashboards will enable real-time tracking and visualisation of gender diversity trends in AI and DS sectors. These dashboards will provide stakeholders with accessible, up-to-date insights into workforce composition, progress, and areas needing improvement. These tools can support evidence-based decision-making and policy development by making data more actionable.

Fourth, integrating ethical AI standards into the data curation and analysis process is vital to ensuring the responsible use of workforce datasets. Establishing governance frameworks prioritising privacy, Transparency, and equity will help align data practices with ethical principles, fostering trust and accountability in the research process.

Finally, cross-regional comparisons  provide valuable insights into global disparities in gender representation and professional involvement in AI and DS. Researchers can identify context-specific challenges and opportunities by analysing differences between developed and developing regions, informing targeted strategies to promote Inclusivity worldwide.

This work can build a fairer, more inclusive, and representative AI and DS workforce by advancing these research directions and bridging the gap between data-driven evidence and policy recommendations. These efforts will address existing inequities and pave the way for a more diverse and innovative future in these rapidly evolving fields.




\section*{Acknowledgments}
This work has been funded by the projects JOWDISAI of the CNRS Institute of Gender, France and SINFONIA of the CNRS AAP program.

\bibliographystyle{abbrvnat}
\bibliography{bibliography}


\end{document}